\documentclass[]{raa}  
\usepackage{graphicx,times}
\usepackage{natbib}
\usepackage{amssymb,amsmath}
\usepackage{multirow}
\usepackage{caption}
\bibpunct{(}{)}{;}{a}{}{,}

\usepackage[a4paper=true,dvipdfm=true,pagebackref=true]{hyperref}
\hypersetup{pdftitle = The title of my PDF, pdfauthor = My name, pdfsubject= The subject, pdfkeywords = keyword1 keyword2 keyword3}
\hypersetup{colorlinks = true, linkcolor = green, anchorcolor = red, citecolor = blue, filecolor = red, pagecolor = red, urlcolor = red}

\begin{document}

   \title{Estimating stellar effective temperatures and detected angular parameters using stochastic particle swarm optimization
}

 \volnopage{ {\bf 2012} Vol.\ {\bf X} No. {\bf XX}, 000--000}
   \setcounter{page}{1}

   \author{Chuan-Xin Zhang\inst{}, Yuan Yuan\inst{}, Hao-Wei Zhang\inst{}, Yong Shuai
      \inst{},  He-Ping Tan\inst{}
   }

   \institute{ School of Energy Science and Engineering, Harbin Institute of Technology, 92, West Dazhi Street, Harbin 150001, PR China; {\it yuanyuan83@hit.edu.cn}\\
\vs \no
   {\small Received 2016 February 20; accepted 2016 April 28}
}

\abstract{Considering features of stellar spectral radiation and survey explorers, we established a computational model for stellar effective temperatures,
detected angular parameters, and gray rates. Using known stellar flux data in some band, we estimated stellar effective temperatures and detected angular parameters
using stochastic particle swarm optimization (SPSO). We first verified the reliability of SPSO, and then determined reasonable parameters that produced highly accurate
estimates under certain gray deviation levels. Finally, we calculated 177,860 stellar effective temperatures and detected angular parameters using the Midcourse Space
Experiment (MSX) catalog data. These derived stellar effective temperatures were accurate when we compared them to known values from literatures. This research made full
use of catalog data and presented an original technique for studying stellar characteristics. It proposed a novel method for calculating stellar effective temperatures and
detected angular parameters, and provided theoretical and practical data support for finding radiation flow information for any band.
}

   \authorrunning{C.-X. Zhang et al. }            
   \titlerunning{Estimating stellar effective temperatures}  
   \maketitle

%
\section{Introduction}           
\label{sect:intro}

In order to obviously (\citep{Ahn+etal+2012})  distinguish targets in sky surveys, appropriate detected bands need to be selected. The bands of existing large scale sky surveys are
listed in Table 1 (\citealt{Neugebauer+etal+1984, Wright+etal+2010, Skrutskie+etal+2006, Ishihara+etal+2010, Egan+etal+2003, Ahn+etal+2012, Zhao+etal+2012,
Kordopatis+etal+2013, Steinmetz+etal+2006, Aihara+etal+2011, Xiang+etal+2015, Yuan+etal+2015, Luo+etal+2015}). The detected bands that already existed are not
comprehensive, and the flux density of full-wave band cannot be obtained. In other words, stellar radiation energy cannot be obtained in some other bands, but
we need energy information of these bands. It is therefore important to develop methods to determine the required radiation energy information from existing
information. Stellar atmospheric parameters (including stellar effective temperatures, surface gravity, and chemical abundances) are important to different stellar
spectral data. Stellar effective temperatures effect luminosities and spectral characteristics, and they are also closely related to stellar physical properties,
chemical compositions, and star evolution (\citealt{Huang+etal+2015}). Thus, if we can obtain stellar effective temperatures and detected angular parameters, we
can reduce the complexity of this problem and derive radiation energy of any band.

Stellar color is determined by its effective temperature. Stellar preliminary information such as effective temperatures can be obtained from the stellar color
or spectral type approximately. We can calculate stellar physical parameters from low-resolution spectra using Indirect Calculation Method, Infrared Flux Method,
Template Matching Method, Neural Network Method, Non-parametric Estimation Method, Color Selection Method, and so forth. Indirect Calculation Method calculated stellar surface effective temperatures using the distance between stars and the Earth, and stellar brightness. \cite{Blackwell+etal+1977} and \cite{Blackwell+etal+1980} used Infrared
Flux Method to calculate stellar effective temperatures and angular diameters. This approach needs a precise sequence of infrared flow and temperature data. However, precise
data regarding these physical properties can only be determined for a limited number of stars, regardless of Indirect Calculation Method and Infrared Flux Method. Only a
small number of stars were precisely measured. \cite{Soubiran+etal+1998} and \cite{Katz+etal+1998} established a stellar spectral template library including 211 stars. They
used Nearest Neighbor Method to calculate stellar spectral radiation fluxes. \cite{Bailer-Jones+etal+2000} calculated analog synthetic spectra and stellar spectral radiation
fluxes using Neural Network Method. \cite{Zhang+etal+2006} proposed fitting and estimating stellar effective temperatures using a polynomial exponential model and Non-parametric
Estimation Method. They first decomposed spectral data using Principal Component Analysis (PCA). They then derived the fitting surface of the polynomial exponential function
that corresponded to its surface temperature using the PCA data. Stellar surface temperatures were calculated using the calculated polynomial logarithmic function finally.
\cite{Engelke+etal+2006} composed a library of stellar spectral templates using spectral data fragments that were observed by spectrometers. They used spectral template technique
proposed by \cite{Cohen+etal+1993}, any segment of spectral radiation flux can be fitted and composed using this spectral template library. However, it is not accurate and fundamental
stellar information (including stellar effective temperatures and detection angles) was not calculated. \cite{Rebassa-Mansergas+etal+2010, Rebassa-Mansergas+etal+2012, Rebassa-Mansergas+etal+2013} searched for white-dwarf-main-sequence
binaries using template matching and Color Selection Method based on the Sloan Digital Sky Survey (SDSS) photometry optical and infrared photometry. \cite{Ren+etal+2014} calculated stellar
temperatures using Template Matching Method, and estimated the distance of DA/M binaries.  They are focused on white dwarf-main sequence binaries.

The above summary shows that existing publications mainly used star catalog data and spectral template technology to calculate stellar atmospheric parameters. Most models
used star catalogs to directly record stellar radiation energy, and simulated bands are limited. We, based on this, applied particle swarm optimization algorithm to these
data to calculate effective stellar temperatures and detected angular parameters considering survey stellar radiation flux data. These two parameters can be used to indirectly
calculate radiation energy of any band which are required. In this work, we investigated impacts of algorithmic parameters, grayscale deviations, and the selection of observed
samples on inversion results. We also analyzed the performance of inversion models and determined the optimal inversion parameters. We compared our results with known data to
verify the accuracy and applicability of the inversion model.

\begin{table}
\bc
\begin{minipage}[]{80mm}
\caption{Sky Survey and Detection Bands.\label{tab1}}\end{minipage}
\setlength{\tabcolsep}{1pt}
\small
 \begin{tabular*}{140mm}{ccccccccccccc}
  \hline\noalign{\smallskip}
Probe name& Detection bands\\
  \hline\noalign{\smallskip}
IRAS, Infrared Astronomical Satellite&12, 25, 60, 100\\
WISE, Wide¨Cfield Infrared Survey Explorer&3.4, 4.6, 12, 22\\
2MASS, Two Micron All Sky Survey&1.25, 1.65, 2.17\\
AKARI (IRIS), Infrared Imaging Surveyor&9, 18\\
MSX, Midcourse Space Experiment&4.25, 4.29, 8.23, 12.13, 14.65, 21.34\\
SDSS, Sloan Digital Sky Survey&0.3551, 0.4686, 0.6166, 0.748, 0.8932\\
LAMSOT, Large sky Area Multi¨CObject fiber Spectroscopic Telescope&0.37\textrm{-}0.9\\
  \noalign{\smallskip}\hline
\end{tabular*}
\ec
\end{table}

\begin{figure}
   \centering
   \includegraphics[width=6.0cm, height=7.0cm, angle=0]{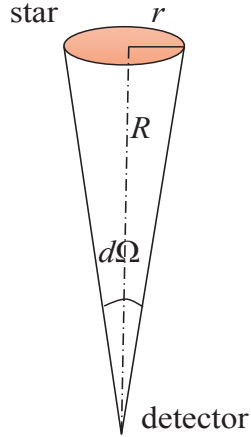}
   \caption{Schematic diagram of the stellar solid angle.}
   \label{Fig1}
   \end{figure}

\section{METHODS}
\label{sect:Obs}

\subsection{Stellar Effective Temperatures Model}

Stars have similar characteristics with blackbodies. However, the presence of gases at different temperatures, pressures, and densities on the stellar
surface means that some bands have significant absorption or emission lines. It is important that we study gases and characteristics of stellar surfaces
in relation to these absorption or emission lines. If stellar atmosphere is supposed to be in a thermodynamic equilibrium state, stellar surface temperatures
can be calculated using the relevant blackbody radiation formula. Assuming stars have similar spectral emissivity and gray characteristics, if these parameters
can meet our requirements regarding detection accuracy, then this assumption is valid.
Planck¡¯s law describes the electromagnetic radiation emitted by a blackbody in thermal equilibrium at a definite temperature, and that the spectral emissive power
of the blackbody in some bands is

\begin{equation}
{E_b}_{\left( {{\lambda _1} - {\lambda _2}} \right)} = \int_{{\lambda _1}}^{{\lambda _2}} {\frac{{{c_1}{\lambda ^{ - 5}}}}{{\exp [{c_2}/(\lambda T)] - 1}}} d\lambda  
\end{equation}

Here, ${E_b}_{\left( {{\lambda _1} - {\lambda _2}} \right)}$ represents the spectral emissive power of the blackbody from $\lambda_1$ to $\lambda_2$ $(W\cdot m^{-2})$; $\lambda$ is the wavelength $(m)$, $T$ is the thermodynamic temperature of the blackbody $(K)$; $c_1$ is the first radiation constant $(c_1=3.7419\times10^{-16} W\cdot m^2)$; and $c_2$ is the second radiation constant $(c_2=1.4388\times10^{-2}m\cdot K)$.

Denote the stellar spectral emissivity as ${\varepsilon _{{\lambda _1} - {\lambda _2}}}$. The band radiation power is

\begin{equation}
{E_{{\lambda _1} - {\lambda _2}}}{\rm{ = }}{\varepsilon _{{\lambda _1} - {\lambda _2}}} \cdot {E_b}_{\left( {{\lambda _1} - {\lambda _2}} \right)}  
\end{equation}

and the band radiation intensity is

\begin{equation}
{I_{{\lambda _1} - {\lambda _2}}}{\rm{ = }}{E_{{\lambda _1} - {\lambda _2}}}/\pi    
\end{equation}

The radius of the effective stellar temperature for the calculation is $r$, and the distance between the stellar surface and the detector receiving surface is $R$. Then, the solid angle between the stellar surface and the detector is

\begin{equation}
d\Omega  = d{A_s}/{R^2} = \pi {r^2}/{R^2}  
\end{equation}

A schematic diagram of the solid angle is shown in Figure 1. The band radiation power of the received detector is

\begin{equation}
{E_{p\left( {{\lambda _1} - {\lambda _2}} \right)}} = d\Omega  \cdot {I_{{\lambda _1} - {\lambda _2}}}   
\end{equation}

The band radiation power of the received detector can be calculated from Equations (1)\textrm{-}(5), and is defined as

\begin{equation}
{E_{p\left( {{\lambda _1} - {\lambda _2}} \right)}} = \frac{{{r^2}}}{{{R^2}}} \cdot {\varepsilon _{{\lambda _1} - {\lambda _2}}} \cdot \int_{{\lambda _1}}^{{\lambda _2}} {\frac{{{c_1}{\lambda ^{ - 5}}}}{{\exp [{c_2}/(\lambda T)] - 1}}} d\lambda    
\end{equation}

Thus, we can define the stellar detected angular parameter as

\begin{equation}
{\xi _{{\lambda _1} - {\lambda _2}}} = {\varepsilon _{{\lambda _1} - {\lambda _2}}} \cdot \left( {{r^2}/{R^2}} \right)   
\end{equation}

Equation (6) can be written as

\begin{equation}
{E_{p\left( {{\lambda _1} - {\lambda _2}} \right)}} = {\xi _{{\lambda _1} - {\lambda _2}}} \cdot \int_{{\lambda _1}}^{{\lambda _2}} {\frac{{{c_1}{\lambda ^{ - 5}}}}{{\exp [{c_2}/(\lambda T)] - 1}}} d\lambda    
\end{equation}

which can be used to determine the band radiation power of the received detector that is above the Earth. In other words, the band radiation power of the received detector is calculated using the stellar effective temperature $(T_{eff})$ and the detected angular parameter $({\xi _{{\lambda _1} - {\lambda _2}}})$,that is,

\begin{equation}
{E_{p\left( {{\lambda _1} - {\lambda _2}} \right)}} = f\left( {{T_{eff}},{\xi _{{\lambda _1} - {\lambda _2}}}} \right)  
\end{equation}

Because the stellar surface contains gases that have different temperatures, pressures, and densities, some bands have significant absorption or emission lines. Therefore, the stellar band radiation power is not exactly the same as a blackbody. To be more specific, there are differences to blackbody radiation in some bands. This deviation is defined as the gray rate,$({\delta _{{\lambda _1} - {\lambda _2}}})$. Eq. (9) can also be written as

\begin{equation}
{E_{{\rm{p}}\left( {{\lambda _1} - {\lambda _2}} \right)}} = f\left( {{T_{eff}},{\xi _{{\lambda _1} - {\lambda _2}}},{\delta _{{\lambda _1} - {\lambda _2}}}} \right)  
\end{equation}

\subsection{Parameters Acquisition for Stellar Effective Temperatures Model}

The fixed band average radiation flux data can be obtained using a satellite detector, but the radiation flux data of the other bands cannot. To solve this problem, we consider the following. We calculate effective stellar temperatures and detected angular parameters using inversion and several fixed band average radiation flux data. We can use this model to get stellar flux data in any band.

Therefore, we set the physical model established in Section 2.1 as the direct problem, and use SPSO to solve the inverse problem. This process is described in Figure 2.

\begin{figure}
   \centering
   \includegraphics[width=14.0cm, height=3.0cm, angle=0]{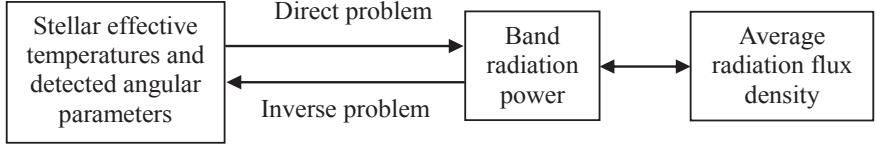}
   \caption{Direct and inverse problems of derived stellar parameters.}
   \label{Fig2}
   \end{figure}

The direct problem can be described as follows. Firstly, we calculate some bands radiation power using known stellar effective temperatures and detected angular parameters. Then, we determine the average radiation flux density data. The corresponding inverse problem calculates some band radiation power using the average radiation flux density data. Then, stellar effective temperatures and detected angular parameters are calculated in an inverse method. The average flux density data are the driving source of the inverse problem. In this paper, these data were obtained from the Midcourse Space Experiment (MSX) catalog.

The MSX was launched by the United States in 1996, and is used for the galactic plane and areas that are not covered by the Infrared Astronomical Satellite (IRAS) infrared observations. It contains an infrared instrument called SPIRIT ¢ó,which is a 35 cm aperture off-axis telescope with a high sensitivity. The data¡°version 2.3 of the MSX point source catalog¡± was used in this paper. The MSX infrared photometric catalog contains the data of 177,860 stars. Estimations of the right ascension, declination, proper motion, and average flux density for six bands of stars are listed in the MSX catalog. The 6 bands are 6.8\textrm{-}10.8 $\mu m$, 4.22\textrm{-}4.36 $\mu m$, 4.24\textrm{-}4.45 $\mu m$, 11.1\textrm{-}13.2 $\mu m$, 13.5\textrm{-}15.9 $\mu m$, and 18.2\textrm{-}25.1 $\mu m$. The flux density of radio sources is

\begin{equation}
{S_v} = \left( {\theta ,\phi } \right)\cos \theta d\Omega 
\end{equation}

where $S_v$ $(W \cdot m^{-2} \cdot Hz^{-1})$ represents the emitted energy.

The brightness of the radio source (also called the intensity) represents the emitted energy per unit frequency interval per unit area per unit time interval per unit solid angle. The integral illumination is obtained by multiplying the average traffic density and the band. That is,

\begin{equation}
{E_{p\left( {{\lambda _1} - {\lambda _2}} \right)}}{\rm{ = }}\int_{{\lambda _1}}^{{\lambda _2}} {{E_\lambda }d\lambda  = } \int_{c/{\lambda _1}}^{c/{\lambda _2}} {{E_\gamma }d\gamma } 
\end{equation}

which corresponds to the frequency radiation power. The MSX catalog table contains estimations of the average flux density for 6 bands of stars. We can determine the radiation power of these 6 bands.

We cannot determine analytical solutions because the equation is nonlinear. We apply numerical methods to the inverse problem. Additionally, it is
difficult to define general numerical methods that produce satisfactory results because the range of effective stellar temperatures is large and detected angular
parameters may range between several orders of magnitude. We used Particle Swarm Optimization (PSO) algorithm that was proposed by \cite{Eberhart+etal+1995} and \cite{Kennedy+etal+2007}. The
algorithm can find a global optimal solution or a good approximate solution. It can jump to another solution space from a local solution space, so the global optimal
solution is determined iteratively. PSO has been studied extensively and applied to many fields. Recently, the SPSO algorithm has been developed. \cite{Yuan+etal+2010} used SPSO
to calculate the inverse problem for atmospheric aerosol size distribution. \cite{Qi+etal+2008,Qi+etal+2011} adopted PSO to analyze the inverse transient radiation in one-dimensional
non-homogeneous participating slabs and retrieve properties of participating media using different spans of radiation signals. \cite{Wang+etal+2011} calculated the absorption
coefficient in a one-dimensional medium and reconstructed the coal fire depth profile using SPSO. We applied the SPSO algorithm to solve the inverse problem of effective stellar
temperatures and detected angular parameters.

\subsection{Analysis of the PSO and the SPSO Algorithms}

In standard PSO, every possible solution is represented as a particle of the population, and each particle has its own position and velocity related to the inverse problem. All particles in the solution space search for the global optimum by pursuing an optimal adaptation that is determined by an objective function.

The mathematical description of PSO is as follows. There are $M$ particles in a D-dimensional search space, and the spatial position of each particle represents a potential solution. The position vector for particle $i$ is ${X_i} = \left( {{x_{i1}},{x_{i2}}, \cdot  \cdot  \cdot {x_{iD}}} \right)$, and the velocity vector is ${V_i} = \left( {{v_{i1}},{v_{i2}}, \cdot  \cdot  \cdot {v_{iD}}} \right)$. The best position that this particle has experienced (i.e., individual best) is ${P_i} = \left( {{p_{i1}},{p_{i2}}, \cdot  \cdot  \cdot {p_{iD}}} \right)$, and is denoted $P_{best}$. The corresponding best position of all the particles (i.e., global best) is denoted by ${P_g} = \left( {{p_{g1}},{p_{g2}}, \cdot  \cdot  \cdot {p_{gD}}} \right)$ and is denoted by $g_{best}$. The particle velocity depends on the personal best and global best, and it is given by

\begin{equation}
{V_i}\left( {t + 1} \right) = w{V_i}\left( t \right) + {c_1}{r_1}\left[ {{P_i}\left( t \right) - {X_i}\left( t \right)} \right] + {c_2}{r_2}[{P_g}\left( t \right) - {X_i}\left( t \right)]
\end{equation}

Here, $t$ is the current iteration, $w$ is the inertia weight, $c_1$ and $c_2$ are constant accelerations, and $r_1$ and $r_2$ are random numbers in $[0,1]$. The new location of $X_i$ is

\begin{equation}
{X_i}\left( {t + 1} \right) = {X_i}\left( t \right){\rm{ + }}{V_i}\left( {t{\rm{ + }}1} \right)
\end{equation}

We set $w=0$, and then get

\begin{equation}
{X_i}\left( {t + 1} \right) = {X_i}\left( t \right) + {c_1}{r_1}\left[ {{P_i}\left( t \right) - {X_i}\left( t \right)} \right] + {c_2}{r_2}[{P_g}\left( t \right) - {X_i}\left( t \right)] 
\end{equation}

This formula reduces the global search capability, but increases the local search capability. So, if $X_j (t)=P_j =P_g$, particle j will ¡°flying¡± at the velocity zero. To improve the global search capability, we conserve the current best position of the swarm $P_g$ and the $j$¡¯s best position $P_j$, then giving a new particle $j$¡¯s position $X_j (t+1)$, and other particles are manipulated according to (15), thus the global search capability enhanced. Because of the particle¡¯s position need to sample from the domain when $X_j (t)=P_j =P_g$, the modified PSO algorithm called stochastic PSO (SPSO).

The standard PSO algorithm may prematurely converge on suboptimal solutions that are not even guaranteed to be local extrema. On the contrast, the stochastic PSO, which is presented based on the analysis of the standard PSO, is more efficiency because of its local search capability according to \cite{Cui+etal+2004}. The SPSO algorithm was compared with the PSO algorithm in the following pages, and then the optimal choice was obtained.

We used the following computational procedure to solve stellar parameters problem.

Step 1: Set the input parameters of the system. The population size was set to 50, but it is adjustable and changeable. The maximum number of iterations was 3000, there were two variables, and we set the acceleration constants to $c_1=1.80$ and $c_2=1.80$. We assumed that the effective temperature was within $1000 \textrm{-} 20000 K$ and that the detected angular parameters was between $1.0 \times 10^{-21}$ and $1.0 \times 10^{-16}$. In this algorithm, we expanded the above parameters to meet the computing requirements. That is, the range of effective temperatures was set to $1.0 \times 10^{2.5} \textrm{-} 1.0 \times 10^{4.5} K$, and range of the detected angular parameters was set to $1.0 \times 10^{-26} \textrm{-} 1.0 \times 10^{-10}$. This determined the boundary of the solution space.

Step 2: Calculate the fitness value of each particle. A particle¡¯s adaptation value is equal to its objective function value. The objective function is

\begin{equation}
Fitnes{s_i}{\rm{ = }}\sqrt {{{\left( {\left( {{E_{ipa}} - {E_{ipb}}} \right)/{E_{ipa}}} \right)}^2}} 
\end{equation}

where Eipa represents the initial value of the inversion, and Eipb represents the value for particle i.

Step 3: Compare the fitness value of each particle with the a priori best, $P_i$. If the fitness is lower than $P_i$, set this value as the current $P_i$, and record the corresponding particle position.

Step 4: Compare the fitness value of each particle with the a priori best $P_g$; if the fitness is lower than $P_g$, set this value as the current $P_g$, and record the corresponding particle position.

Step 5: Generate the new particle, and update the velocity and position of the other particles using Equations (13) and (15). If $X_j (t)=P_j=P_g$, the $j$¡¯s position are generated randomly.

Step 6: Check the stopping criteria. If we have reached the pre-set maximum number of generations $(1.0 \times 10^{-11})$ or if there was no improvement to the best solution after a given number of iterations (3000), then the process is terminated. Otherwise, we increment the iteration index $(t=t+1)$ and go back to Step 2.

We compared the performances of the PSO and SPSO algorithms. For standard PSO, $w=1.0$, and for SPSO $w=0.0$. There were two termination criteria: (1) when the iteration accuracy was below a level fixed of $10^{-10}$ and (2) when we reached more than 3000 generations. We used five particles for both algorithms. The results are compared in Figure 3. The SPSO algorithm converged much faster than the standard PSO algorithm. Moreover, the SPSO algorithm found better values than the standard PSO algorithm with a smaller number of generations. Therefore, we applied SPSO to solve the stellar parameters problem.

\begin{figure}
\begin{minipage}[t]{0.5\linewidth}
   \centering
   \includegraphics[width=7.0cm, angle=0]{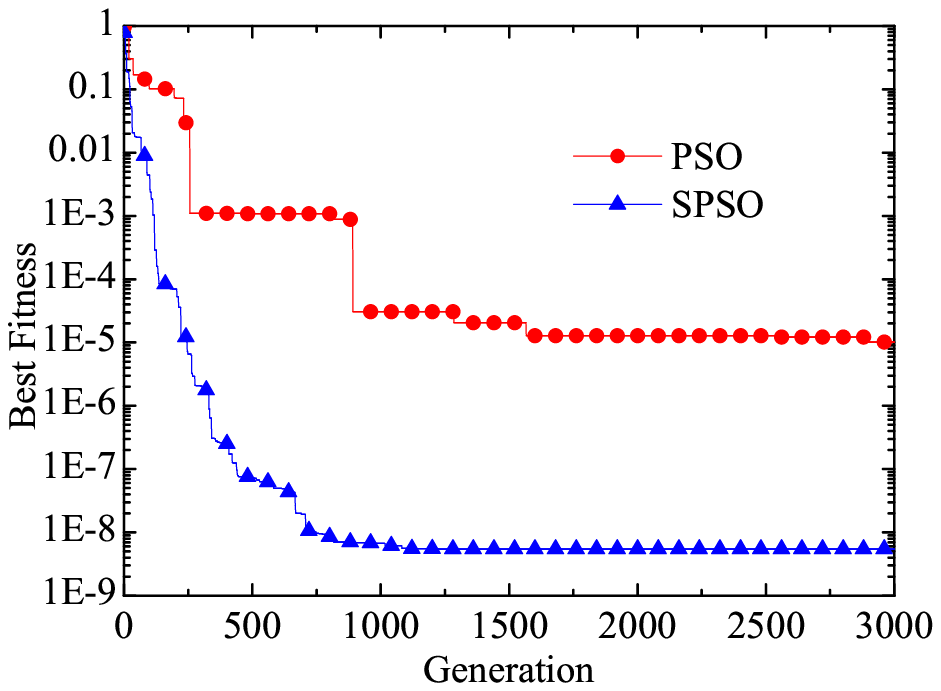}
   \caption{Comparison of the values found by PSO and SPSO.}
   \label{Fig3}
   \end{minipage}
\begin{minipage}[t]{0.5\linewidth}
   \centering
   \includegraphics[width=7.0cm, angle=0]{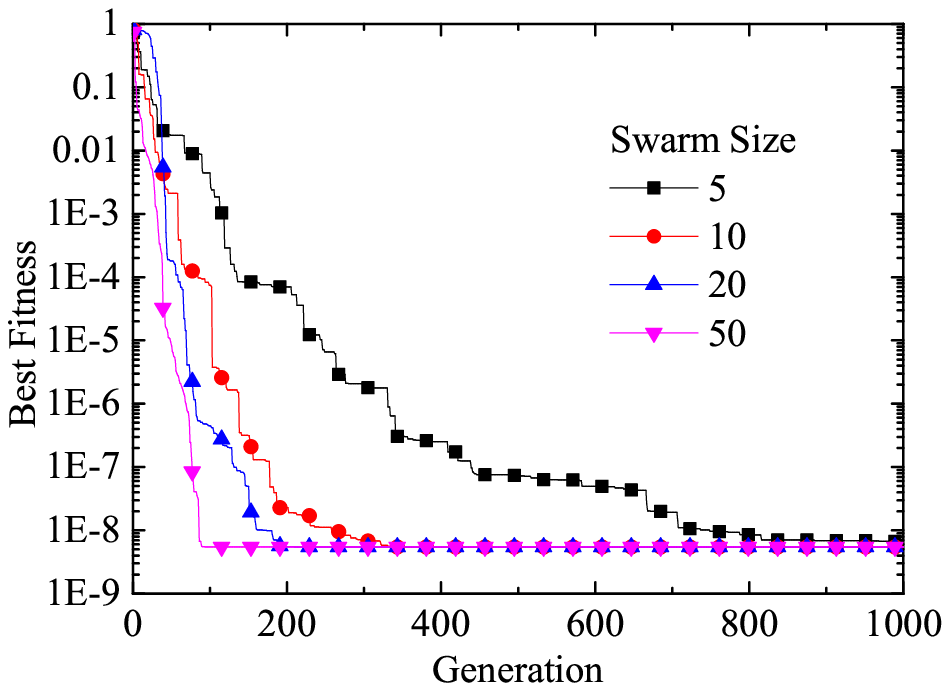}
   \caption{Comparison of the best fitness values for different swarm sizes.}
   \label{Fig4}
   \end{minipage}
\end{figure}

\section{NUMERICAL EXPERIMENTS AND RESULTS}
\label{sect:results}

Stellar effective temperatures and detected angular parameters can be calculated using the above SPSO algorithm if the stellar band radiation power is known. The accuracy of results is related to initial values of parameters in the inversion algorithm, the selected inversion bands, and the applicable inversion temperature range. To verify the accuracy and stability of the SPSO algorithm for calculating stellar effective temperatures and detected angular parameters, we analyzed the following examples and studied the three factors separately.

The six band radiation power is different if effective temperatures and detected angular parameters are different. Table 2 shows the band radiation power data for different initial temperatures.

\begin{table}
\bc
\begin{minipage}[]{120mm}
\caption[]{Six Bands Radiation Power Data at Different Effective Temperatures.\label{tab2}}\end{minipage}
\setlength{\tabcolsep}{1pt}
\small
 \begin{tabular}{ccccccccccccc}
  \hline\noalign{\smallskip}
\multicolumn{2}{c}{Initial parameters} & \multicolumn{6}{c}{Band radiation power $(10^{-16}W \cdot m^{-2}) $}\\
\hline
Detected angular~~ & Effective~~  & 6.8\textrm{-}10.8~~& 4.22\textrm{-}4.36~~& 4.24\textrm{-}4.45~~& 11.1\textrm{-}13.2~~& 13.5\textrm{-}15.9~~& 18.2\textrm{-}25.1~~\\
parameters& temp.(K)~~ & ($\mu m$)& ($\mu m$)& ($\mu m$)& ($\mu m$)& ($\mu m$)& ($\mu m$)\\
  \hline\noalign{\smallskip}
                & 1000& 15.039& 2.611& 3.841& 2.661& 1.601&	1.234\\
 $2 \times 10^{-19}$& 5000& 172.210& 75.497& 108.222& 22.736& 12.359& 82.945\\
                & 10000& 377.966& 181.093& 258.973& 48.369& 26.005& 17.184\\
  \noalign{\smallskip}\hline
\end{tabular}
\ec
\end{table}

Case 1. Number of particles

The number of particles in the SPSO algorithm has an important effect on the inversion efficiency, and it is directly related to the accuracy. Thus, we must determine the optimal number of particles.

We set the number of particles to 5, 10, 20, and 50 and calculated the results in Figure 4. An increase of the particle number corresponds to a decrease in the number of generations required for the algorithm to converge. When the convergence value was set to $1.0 \times 10^{-8}$, 749 generations were required when there were five particles, whereas only 86 generations were required when there were 50 particles.

When calculating effective temperatures and detected angular parameters, nonlinearities mean that more particles significantly increase the computation time, but do not help the convergence of the residual. Therefore, considering the computation time and the accuracy, we selected 50 particles.

Case 2. Selection of the inversion band

From the six bands of the MSX catalog, three bands of radiation power data were used as initial values for the inversion. The six combinations (Plan A, B, C, D, E, and F) are shown in Table 3.

\begin{table}
\bc
\begin{minipage}[]{70mm}
\caption{Different Combinations Plans.\label{tab3}}\end{minipage}\\
\setlength{\tabcolsep}{1pt}
\small
 \begin{tabular}{ccccccccccccc}
  \hline\noalign{\smallskip}
~~Different Plans~~& ~~Bands Combinations ($\mu m$)~~\\
  \hline\noalign{\smallskip}
Plan A&~~6.8\textrm{-}10.8,~~ 4.22\textrm{-}4.36,~~ 4.24\textrm{-}4.45~~\\
Plan B&~~6.8\textrm{-}10.8,~~ 4.24\textrm{-}4.45,~~ 13.5\textrm{-}15.9~~\\
Plan C&~~6.8\textrm{-}10.8,~~ 4.24\textrm{-}4.45,~~ 18.2\textrm{-}25.1~~\\
Plan D&~~6.8\textrm{-}10.8,~~ 13.5\textrm{-}15.9,~~ 18.2\textrm{-}25.1~~\\
Plan E&~~4.22\textrm{-}4.36,~~ 4.24\textrm{-}4.45,~~ 11.1\textrm{-}13.2~~\\
Plan F&~~11.1\textrm{-}13.2,~~ 13.5\textrm{-}15.9,~~ 18.2\textrm{-}25.1~~\\
  \noalign{\smallskip}\hline
\end{tabular}
\ec
\end{table}

Stellar effective temperatures and detected angular parameters depend on the inversion plans. This reflects the different properties of stellar information.

To demonstrate the effect of the gray rate $(\delta)$ on the inverse parameters, we added random standard deviations to the exact parameters computed from the direct problem. That is,

\begin{equation}
{Y_{mea}} = {Y_{exact}} + \sigma  \cdot \varsigma 
\end{equation}

where $\zeta$ is a normally distributed random variable with zero mean and unit standard deviation. The standard deviation of the measured ${\emph{E} _{{\lambda _1} - {\lambda _2}}}$, for a $\gamma$ measured error at 99\% confidence, is

\begin{equation}
\sigma {\rm{ = }}\left( {{Y_{{\rm{exact}}}} \times \gamma {\rm{\% }}} \right)/2.576 
\end{equation}

For comparison, the relative error is

\begin{equation}
{\varepsilon _{{\rm{rel}}}}{\rm{ = 100}} \times \left( {{Y_{{\rm{est}}}}{\rm{ - }}{Y_{{\rm{exact}}}}} \right)/{Y_{{\rm{exact}}}}
\end{equation}

The stellar effective temperature was set to 5000 K, and the detected angular parameter was set to $2.0 \times 10^{-19}$. We then calculated the six bands radiation flux data. Normally distributed deviations were added to the radiation flux data, and then the data with no deviations were used to solve the inverse calculations of the effective temperature and detected angular parameter. The data for the six plans are shown in Table 3. The variations to the effective temperature and detected angular parameter with gray deviations are shown in Figures 5 and 6.

As shown in Figure 5 and 6, the relative errors of the six plans were very close to 0 when there were no gray deviations. The estimated parameters are very consistent with the true values. This shows that each plan satisfies the requirement that there is no gray deviations. When there were gray deviations, an increase in gray deviations corresponded to an increase in the relative error of the effective temperature and detected angular parameter. The relative errors for Plans A, C, and D increased more rapidly and significantly than the gray rate. When the gray rate increased by 30\%, the effective temperature for Plans A and C increased by 157.5\% and 112.7\%, while the detected angular parameter of Plan D increased by 141.4\%. This shows that there are strong absorption or emission lines when using these three plans, and that they produce the effective temperature and detected angular parameter values that represent deviations from the initial spectral radiation power data.

The inversion results for Plans B, E, and F were relatively good, as shown by the detailed errors in Figure 7 and 8. The inversion results for Plan B were the closest to the true values. When the gray rate increased by 30\%, the relative errors in the effective temperature and detected angular parameter increased by 5.4\% and 5.1\%. Therefore, the relative errors of the inversion results were much smaller than the gray deviations, indicating that the inversion results for Plan B were the best.

In the above results, the effective temperature was set to 5000 K. Considering the large range of the effective temperature, different effective temperatures correspond to different bands of radiation power. We must verify the accuracy of the results for different effective temperatures.

Next, we verified the accuracies of stellar effective temperatures and detected angular parameters for different radiation power data obtained from different effective temperatures (1000 K and 10000 K)

When set the effective temperature to 1000 K or 10000 K, and used the obtained band radiation power in the inverse problem. The results are shown in Tables 4 and 5. When the effective temperature was 1000 K and the gray rate increased by 30\%, the relative errors in the effective temperature and detected angular parameter only increased by 1.4805\% and 2.329\%. When the effective temperature was 10000 K, the relative errors in the effective temperature and detected angular parameter only increased by 8.769\% and 7.820\%. The relative errors in the inversion parameters were much smaller than the gray rate, and the estimates were close to the initial data. Therefore, the inversion requirement was satisfied.

\begin{table}
\bc
\begin{minipage}[]{150mm}
\caption[]{Relative Errors for Different Gray Deviations When $T_{eff}=1000K$ using SPSO.\label{tab4}}\end{minipage}
\setlength{\tabcolsep}{1pt}
\small
 \begin{tabular}{ccccccccccccc}
  \hline\noalign{\smallskip}
\multirow{2}{*}{~~Parameter~~}
&\multirow{2}{*}{~~True value~~}
&\multicolumn{2}{c}{$\gamma=0$}& \multicolumn{2}{c}{$\gamma=5$}&  \multicolumn{2}{c}{$\gamma=15$}&  \multicolumn{2}{c}{$\gamma=30$}\\
\cline{3-10}
& & ~~SPSO& $\varepsilon_{rel}\%$~~& ~~SPSO& $\varepsilon_{rel}\%$~~& ~~SPSO& $\varepsilon_{rel}\%$~~& ~~SPSO& $\varepsilon_{rel}\%$~~\\
\hline\noalign{\smallskip}
 $T_{eff}$& ~~1000.0~~& ~~1000.0~~& ~~0.000~~& ~~999.3~~& ~~0.069~~& ~~1001.4~~& ~~0.142~~& ~~1014.8~~& ~~1.4805~~\\
 ~~$\zeta \times 10^{-19}$~~& 2.0000& 2.0000& 0.000& 2.0163& 0.813& 2.0211& 1.057& 1.9534& 2.329\\
  \noalign{\smallskip}\hline
\end{tabular}
\ec
\end{table}

\begin{table}
\bc
\begin{minipage}[]{150mm}
\caption[]{Relative Errors for Different Gray Deviations When $T_{eff}=10000K$ using SPSO.\label{tab5}}\end{minipage}
\setlength{\tabcolsep}{1pt}
\small
 \begin{tabular}{ccccccccccccc}
  \hline\noalign{\smallskip}
\multirow{2}{*}{~~Parameter~~}
&\multirow{2}{*}{~~True value~~}
&\multicolumn{2}{c}{$\gamma=0$}& \multicolumn{2}{c}{$\gamma=5$}&  \multicolumn{2}{c}{$\gamma=15$}&  \multicolumn{2}{c}{$\gamma=30$}\\
\cline{3-10}
& & ~~SPSO& $\varepsilon_{rel}\%$~~& ~~SPSO& $\varepsilon_{rel}\%$~~& ~~SPSO& $\varepsilon_{rel}\%$~~& ~~SPSO& $\varepsilon_{rel}\%$~~\\
\hline\noalign{\smallskip}
  $T_{eff}$~~&~~10000.0~~& ~~10000.0~~ & ~~0.000~~& ~~9724.2~~ & ~~2.758~~ & ~~9633.0~~ & ~~3.670~~ & ~~10876.9& 8.769~~\\
  ~~$\zeta \times 10^{-19}$~~& 2.0000& 2.0000& 0.000& 2.0763& 3.813& 2.1138& 5.688& 1.8436& 7.820\\
  \noalign{\smallskip}\hline
\end{tabular}
\ec
\end{table}

The relative errors in the estimates obtained by the inverse calculations are less than the gray deviations of the raw data for different effective temperatures. Plan B can be used to calculate the raw data without gray deviations, and to calculate the original data with gray deviations.

\begin{figure}
\begin{minipage}[t]{0.5\linewidth}
   \centering
   \includegraphics[width=7.0cm, angle=0]{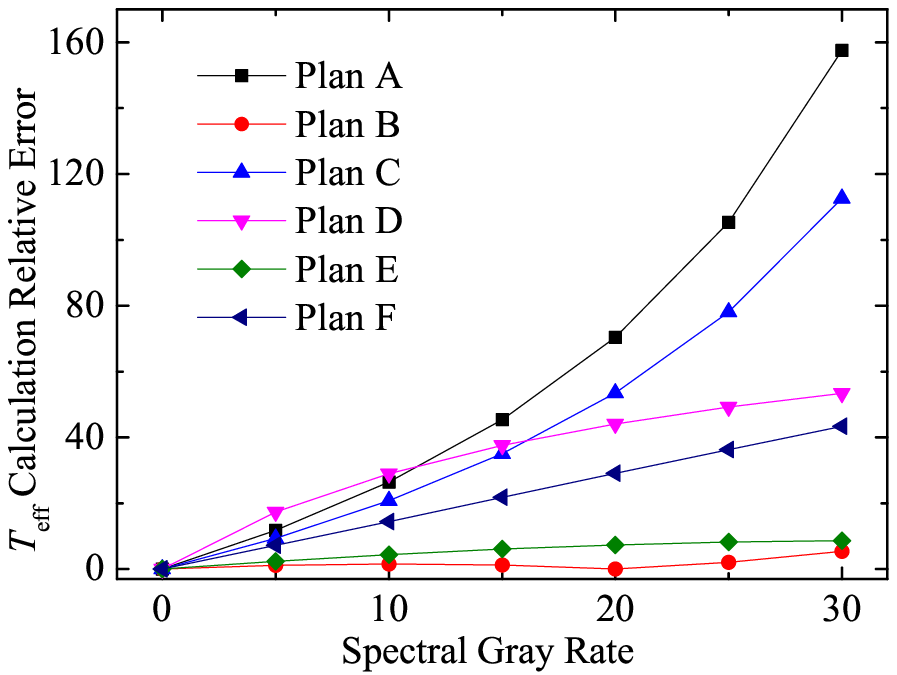}
   \caption{$T_{eff}$ relative error compared with the gray rate for the six plans.}
   \label{Fig5}
   \end{minipage}
\begin{minipage}[t]{0.5\linewidth}
   \centering
   \includegraphics[width=7.0cm, angle=0]{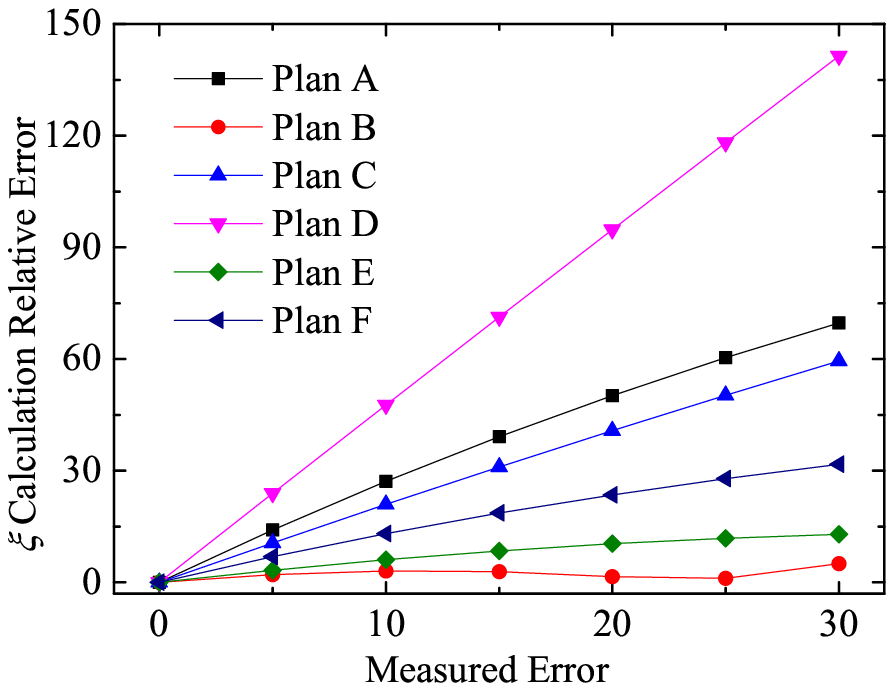}
   \caption{Relative error compared with the gray rate for the six plans.}
   \label{Fig6}
   \end{minipage}
\end{figure}

\begin{figure}
\begin{minipage}[t]{0.5\linewidth}
   \centering
   \includegraphics[width=7.0cm, angle=0]{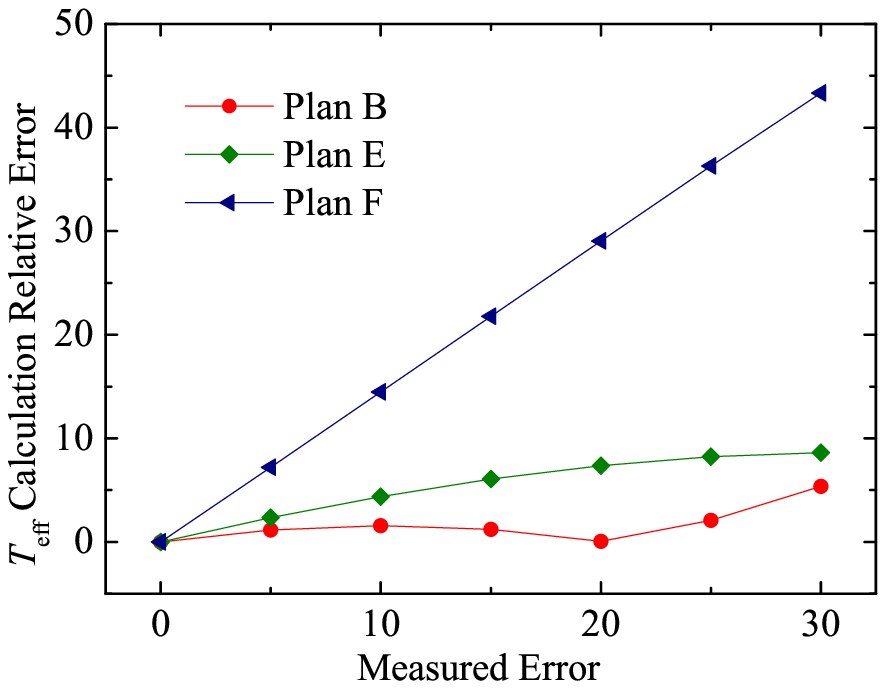}
   \caption{$T_{eff}$ relative error compared with the gray rate for the three plans.}
   \label{Fig7}
   \end{minipage}
\begin{minipage}[t]{0.5\linewidth}
   \centering
   \includegraphics[width=7.0cm, angle=0]{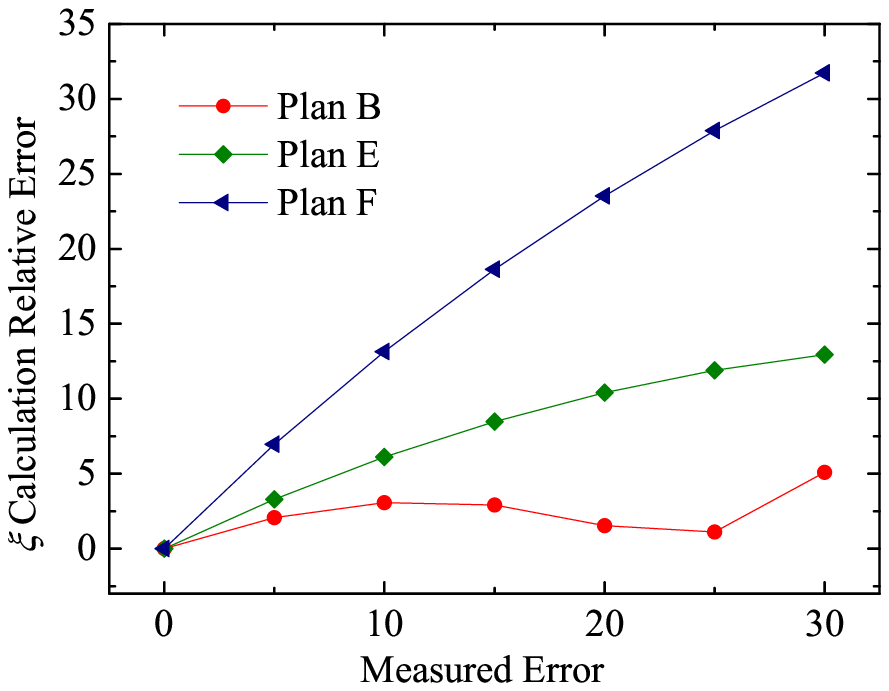}
   \caption{Relative error compared with the gray rate for the three plans.}
   \label{Fig8}
   \end{minipage}
\end{figure}

Case 3.The impact of repeat calculations on the inversion results

Repeat calculations affect the inversion results. As shown in Table 6, when the effective temperature was 5000 K, the detected angular parameter was $2.0 \times 10^{-19}$, and the gray deviation was 30\%, the relative errors changed with repeated calculations. The relative error of the effective temperature was stable at 5.369\%, and the relative error of the detected angular parameter was stable at 5.098\%. This shows that the method can stably calculate the effective temperature and the detected angular parameter.

\begin{table}
\bc
\begin{minipage}[]{150mm}
\caption[]{Results for Different Numbers of Calculations, Demonstrating the Stability of the Inversion.\label{tab6}}\end{minipage}
\setlength{\tabcolsep}{1pt}
\small
 \begin{tabular}{ccccccccccccc}
  \hline\noalign{\smallskip}
~~Calculation~~& Effective temperatures~~& Error~~&  ~~Detected angular~~& ~~ Relative~~ \\
numbers& (K)&  (\%)&  parameters($10^{-19}$)&   ~~error (\%)\\
  \hline\noalign{\smallskip}
1& 5268.4705&	5.369&	1.8980343&	5.098\\
2& 5268.4714&	5.369&	1.8980339&	5.098\\
3& 5268.4711&	5.369&	1.8980340&	5.098\\
4& 5268.4711&	5.369&	1.8980340&	5.098\\
5& 5268.4706&	5.369&	1.8980343&	5.098\\
6& 5268.4713&	5.369&	1.8980339&	5.098\\
7& 5268.4694&	5.369&	1.8980348&	5.098\\
8& 5268.4693&	5.369&	1.8980348&	5.098\\
9& 5268.4702&	5.369&	1.8980344&	5.098\\
10&	5268.4705&	5.369&	1.8980343&	5.098\\

  \noalign{\smallskip}\hline
\end{tabular}
\ec
\end{table}

Our analysis below is based on the above results. Although the maximum radiation flux was for Plan A, the radiation wavelength interval was too small. This resulted in a large inversion error. There were long bands of 18.2\textrm{-}25.1 $\mu m$ for Plans C and D, which also caused large inversion errors. However, Plan B did not have these problems, and produced the best inversion results. It used larger band intervals of 6.8\textrm{-}10.8 $\mu m$, 4.24\textrm{-}4.45 $\mu m$ and 13.5\textrm{-}15.9 $\mu m$, which have big radiation fluxes. So we used Plan B as the inverse solution for the stellar flux data of the MSX catalog.

\section{DATA COMPARISON AND ANALYSIS}
\label{sect:discussion}

The results of the above numerical experiments demonstrated that the model and SPSO algorithm produced highly accurate and stable calculations of stellar effective temperatures and detected angular parameters. We determined the accuracy using an error analysis of different temperatures and different band combinations, and the stability by analyzing repeated calculations. This model and algorithm can solve the stellar flux data inversion problem for effective temperatures between 1000 and 20000 K, detected angular parameters between $1.0 \times 10^{-21} \textrm{-} 1.0 \times 10^{-16}$, and the gray rate below 30\%. When calculating the radiation flux data for the MSX catalog, we used the parameter settings and algorithm suggested by the above analysis.

We calculated the stellar effective temperatures and detected angular parameters for the MSX catalog using the average radiation flux data for each band based on the proposed algorithm. Figure 9 compares our results with 336 true values known in literatures (\citealt{Alves-Brito+etal+2010, Bergemann+etal+2008, Bihain+etal+2004, Burris+etal+2000, carney+etal+2003, Charbonnel+etal+2005, Fulbright+etal+2003, Gratton+eta+2000, Hansen+etal+2012, Ishigaki+etal+2010, Gratton+etal+2003, Ishigaki+etal+2012, Jonsell+etal+2005, Ishigaki+etal+2013, Mishenina+etal+2001, Reddy+etal+2006, Roederer+etal+2008, Saito+etal+2009, Simmerer+etal+2004, Takada-Hidai+etal+2002, Takeda+etal+2005, VanEck+etal+2003, Yong+etal+2003, Fulbright+etal+2000, Hog+etal+2000, Monet+etal+2003, Marshall+etal+2007, Kiraga+etal+2012}). The estimated effective temperatures are consistent with previously reported data. The errors are shown at the bottom in Figure 9. Most errors were less than 10\%, except the individual stars. These inversion results are quite reliable.

\begin{figure}
   \centering
   \includegraphics[width=9.0cm, angle=0]{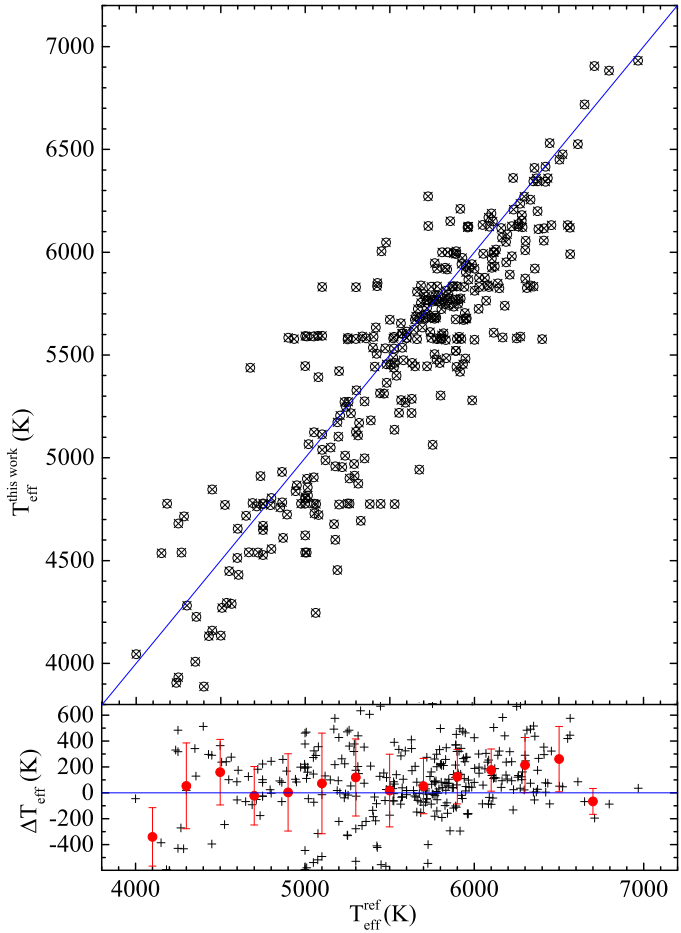}
   \caption{Comparison of effective temperatures in literatures and estimated values in this work. The differences, $\Delta {T_{{\rm{eff}}}}{\rm{ = }}\left( {T_{{\rm{eff}}}^{{\rm{r}}e{\rm{f}}}{\rm{  -  }}T_{{\rm{eff}}}^{{\rm{this work}}}} \right)$ are plotted at the bottom (with red dots and error bars representing the means and standard deviations of differences in the individual temperature bins). }
   \label{Fig9}
   \end{figure}

\begin{figure}
\begin{minipage}[t]{0.5\linewidth}
   \centering
   \includegraphics[width=7.0cm, angle=0]{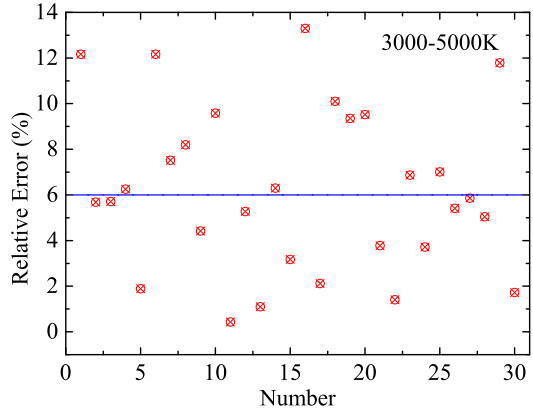}
   \caption{Relative errors for different $T_{eff}$ between 3000\textrm{-}5000 K.}
   \label{Fig10}
   \end{minipage}
\begin{minipage}[t]{0.5\linewidth}
   \centering
   \includegraphics[width=7.0cm, angle=0]{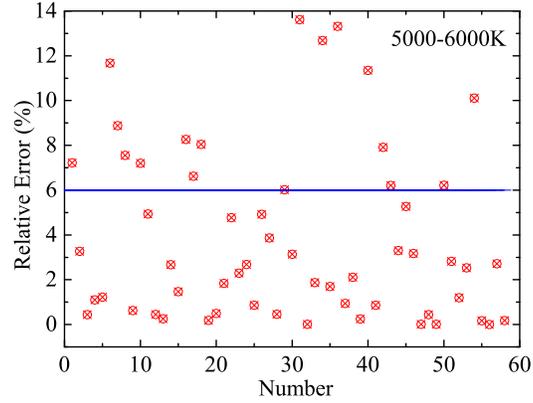}
   \caption{Relative errors for different $T_{eff}$ between 5000\textrm{-}6000 K.}
   \label{Fig11}
   \end{minipage}
\end{figure}

\begin{figure}
   \centering
   \includegraphics[width=7.0cm, angle=0]{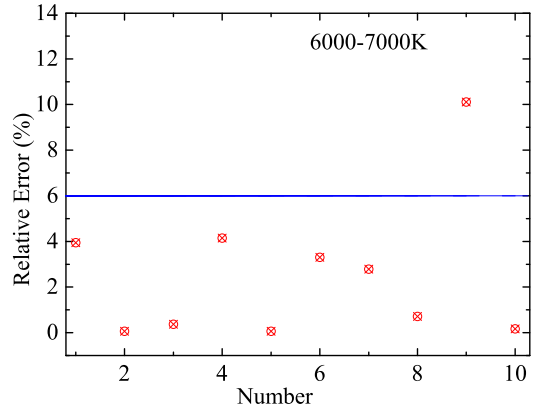}
   \caption{Relative errors for different $T_{eff}$ between 6000\textrm{-}7000K. }
   \label{Fig12}
   \end{figure}

\begin{figure}
   \centering
   \includegraphics[width=7.0cm, angle=0]{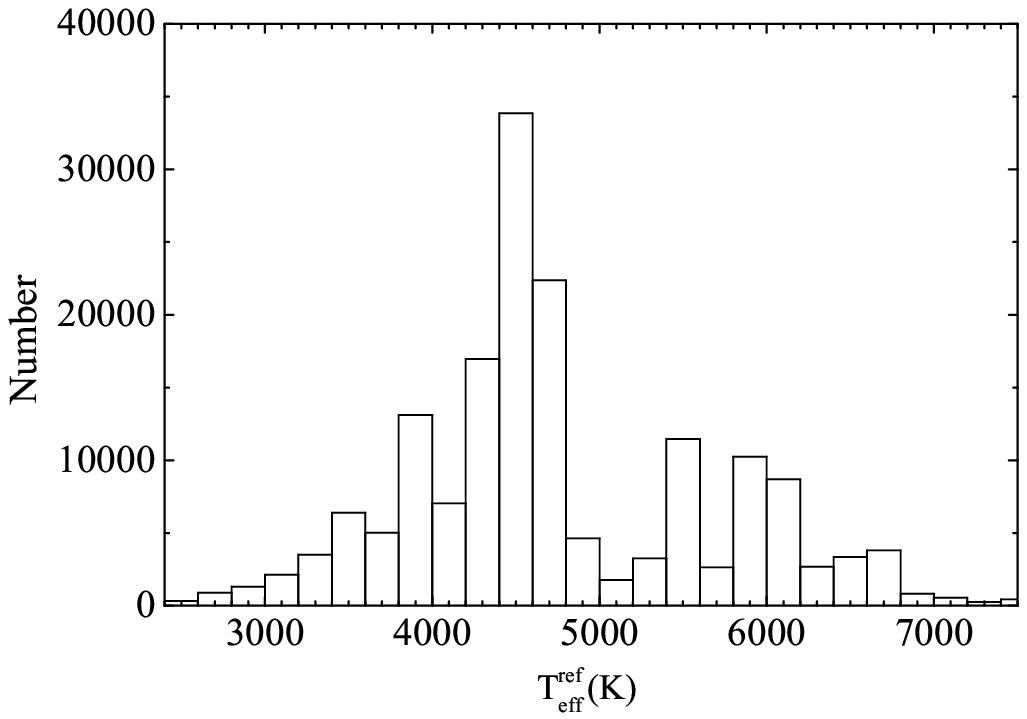}
   \caption{Histogram distribution of stellar effective temperatures in MSX catalog. }
   \label{Fig13}
   \end{figure}

Stellar effective temperatures calculated using the MSX catalog are mostly concentrated in 3000\textrm{-}7000 K. The relative errors are different within this temperature range. The relative errors for three temperature ranges (3000\textrm{-}5000 K, 5000\textrm{-}6000 K, and 6000\textrm{-} 7000 K) are shown in Figure 10, 11 and 12. We can use these results to determine the best inversion temperature range of this computational model. The errors were uniformly distributed around 6\% for effective temperatures of 3000\textrm{-}5000 K, whereas they were mostly less than 6\% for effective temperatures of 5000\textrm{-}6000 K. The overwhelming majority of errors were less than 6\% for effective temperatures of 6000\textrm{-}7000 K. Therefore, this model is most suitable for high temperatures between 6000\textrm{-}7000 K. This conclusion is consistent with the theory. According to Plank¡¯s law, higher effective temperatures correspond to larger radiation fluxes. The impact on the high temperature section is smaller than the low temperature section at the same gray rate.

The histogram distribution of $T_{eff}$ in MSX catalog is plotted as Fig. 13. As we can see, the stellar effective temperatures mainly ranges from 3000K to 7000K, and temperature from 4000K to 6000K is a hot region. The errors in this ranges are acceptable. The means of differences in the individual temperature bins from 4300K to 6000K close to 0K, and the errors of temperatures are within 200K. The model and method apply to temperatures calculation of the MSX catalog.

\section{SUMMARY}
\label{sect:summary}
We used features of stellar spectral radiation and survey explorers to establish a computational model of stellar effective temperatures, detected angular parameters, and gray rates. We applied known stellar flux data in some band to determine stellar effective temperatures and detected angular parameters using SPSO. We first verified the reliability of the SPSO algorithm, and then found reasonable parameters that produced accurate estimates under certain gray deviation levels. Finally, we calculated 177,860 stellar effective temperatures and detected angular parameters using the MSX catalog data. We found that the estimated stellar effective temperatures were very accurate when compared with stellar effective temperatures that are known in literatures.
We selected bands of 6.8\textrm{-}10.8 $\mu m$, 4.24\textrm{-}4.45 $\mu m$, and 13.5\textrm{-}15.9 $\mu m$ in our inversion. This was very accurate. The gray deviation has a smaller impact on the inversion results if the bands are close to the short wave direction for temperatures of 6000\textrm{-}7000 K. This work makes full use of catalog data and presents a new way of studying stellar characteristics. It proposes a novel way of calculating stellar effective temperatures and detected angular parameters.

\normalem
\begin{acknowledgements}

This work was supported by the National Natural Science Foundation of China (Grant No. 51327803 \& 51406041), and
the Fundamental Research Funds for the Central Universities (Grant No. HIT. NSRIF. 2014090). A very special acknowledgement
is made to the editors and referees who made important comments to improve this paper.

\end{acknowledgements}


\end{document}